\documentclass[conference]{IEEEtran}

\usepackage{amsmath,amssymb,amsfonts}
\usepackage[linesnumbered,ruled,vlined]{algorithm2e}
\usepackage{graphicx}
\usepackage{tablefootnote}
\usepackage{textcomp}
\usepackage{listings}
\usepackage{fancyhdr}
\usepackage{comment}
\usepackage{mathptmx} 
\usepackage{soul}
\usepackage{multirow}
\usepackage{booktabs}
\usepackage[normalem]{ulem}
\usepackage[sort,nocompress]{cite}
\usepackage[utf8]{inputenc}
\usepackage{amssymb}
\usepackage{pifont}
\usepackage{tablefootnote}
\usepackage[T1]{fontenc}
\usepackage[utf8]{inputenc}  
\usepackage[T1]{fontenc}     
\usepackage[hyphens]{url}
\usepackage[final]{microtype}
\usepackage[table,xcdraw,dvipsnames]{xcolor}
\usepackage[keeplastbox]{flushend}

\usepackage{enumitem}
\setlist[enumerate]{noitemsep}

\usepackage[bookmarks=true,breaklinks=true,letterpaper=true,colorlinks,linkcolor=blue,citecolor=blue,urlcolor=blue]{hyperref}

\newcommand{\insertFigure}[2]{
    \begin{figure}[!t]
        \centering
        \includegraphics[width=\linewidth]{figures/#1.pdf}
	\vspace{-6mm}
        \caption{ #2}
	\vspace{0mm}
        \label{fig:#1}
    \end{figure}
}

\newcommand{\insertFigureScaled}[3]{
    \begin{figure}[t]
        \centering
        \includegraphics[width=#3\linewidth]{figures/#1.pdf}
	\vspace{-4mm}
        \caption{ #2}
	\vspace{0mm}
        \label{fig:#1}
    \end{figure}
}

\newcommand{\DataflowNamenospace}[0]{\textsc{Score}}
\newcommand{\DataflowName}[0]{\textsc{Score}~}
\newcommand{\DataflowNameTitle}[0]{\textsc{Score}}
\newcommand{\DataflowNameexp}[0]{\underline{S}cheduler for \underline{C}omplex Inter-\underline{O}peration \underline{Re}use}

\newcommand{\SpadNamenospace}[0]{\textsc{Chord}}
\newcommand{\SpadName}[0]{\textsc{Chord}~}
\newcommand{\PolicyA}[0]{\textsc{Prelude}~}

\newcommand{\PolicyAnospace}[0]{\textsc{Prelude}}
\newcommand{\PolicyB}[0]{\textsc{Riff}~}

\newcommand{\PolicyBnospace}[0]{\textsc{Riff}}

\newcommand{\SpadNameTitle}[0]{{\textsc{Chord}}}
\newcommand{\SpadNameexp}[0]{\underline{C}apacity \underline{H}andling via \underline{O}perand-level \underline{R}euse of \underline{D}ata}
\newcommand{\AccelNamenospace}[0]{\textsc{Cello}}
\newcommand{\AccelName}[0]{\textsc{Cello}~}
\newcommand{\AccelNameexp}[0]{Ac\underline{CEL}erator for applications with \underline{L}ow-intensity \underline{O}perations.}

\newcommand{\insertWideFigure}[2]{

    \begin{figure*}[ht!]
        \centering
        \includegraphics[width=\textwidth]{figures/#1.pdf}
	\vspace{-6mm}
        \caption{#2}
	\vspace{0mm}
        \label{fig:#1}
    \end{figure*}

}

\newcommand{\revision}[1]{{\color{ForestGreen}#1}}

\newcommand{\insertWideFigureScaled}[3]{

    \begin{figure*}[ht!]
        \centering
        \includegraphics[width=#3\textwidth]{figures/#1}
	\vspace{-3mm}
        \caption{#2}
	\vspace{0mm}
        \label{fig:#1}
    \end{figure*}

}

\renewcommand{\revision}[1]{{#1}}

\newcommand{\squishlist}{
 \begin{list}{$\bullet$}
  { \setlength{\itemsep}{0pt}
     \setlength{\parsep}{3pt}
     \setlength{\topsep}{3pt}
     \setlength{\partopsep}{0pt}
     \setlength{\leftmargin}{1.5em}
     \setlength{\labelwidth}{1em}
     \setlength{\labelsep}{0.5em} } }

\newcommand{\squishlisttwo}{
 \begin{list}{$\arabic$}
  { \setlength{\itemsep}{0pt}
     \setlength{\parsep}{0pt}
    \setlength{\topsep}{0pt}
    \setlength{\partopsep}{0pt}
    \setlength{\leftmargin}{2em}
    \setlength{\labelwidth}{1.5em}
    \setlength{\labelsep}{0.5em} } }

\newcommand{\squishend}{
  \end{list}  }

\newcommand{\greencheck}{{\color{teal}\checkmark}}

\newcommand{\redcheck}{{\color{red}\xmark}}

\newcommand{\TODO}[1]{\textcolor{blue}{TODO::: #1}}
\newcommand{\RG}[1]{{\color{orange}\bfseries [Raveesh::: #1]}}
\newcommand{\MP}[1]{{\color{olive}\bfseries [Michael::: #1]}}
\newcommand{\SR}[1]{{\color{purple}\bfseries [Siva::: #1]}}
\newcommand{\TK}[1]{{\color{red}\bfseries [Tushar::: #1]}}

\newcommand{\GEMM}[0]{skewed GEMMs~}

\newcommand{\xmark}{\ding{55}}%
\newcommand{\reviewme}[1]{{{#1}}}

\usepackage{url}

\title{\AccelNamenospace: Co-designing Schedule and Hybrid Implicit/Explicit Buffer for Complex Tensor Reuse
}

\author{\IEEEauthorblockN{Raveesh Garg}
\IEEEauthorblockA{
\textit{Georgia Institute of Technology}\\
raveesh.g@gatech.edu}
\and
\IEEEauthorblockN{Michael Pellauer}
\IEEEauthorblockA{
\textit{NVIDIA}\\
mpellauer@nvidia.com}
\and
\IEEEauthorblockN{Sivasankaran Rajamanickam}
\IEEEauthorblockA{
\textit{Sandia National Laboratories}\\
srajama@sandia.gov}
\and
\IEEEauthorblockN{Tushar Krishna}
\IEEEauthorblockA{
\textit{Georgia Institute of Technology}\\
tushar@ece.gatech.edu}
}

\begin{document}
\maketitle

\vspace{-3mm}

\begin{abstract}

Tensor algebra accelerators have been gaining popularity for running high-performance computing (HPC) workloads. 
Identifying optimal schedules for individual tensor operations and designing hardware to run these schedules is an active area of research. 
Unfortunately, operators in HPC workloads such as Conjugate Gradient often have operators with skewed shapes, fundamentally limiting the reuse any schedule can leverage. 
Moreover, the operators form a complex DAG of dependencies, making it challenging to apply simple fusion/pipelining techniques to extract inter-operation reuse.
To address these challenges, 
this work proposes an accelerator \AccelNamenospace. \AccelName uses a novel on-chip buffer mechanism called \SpadName co-designed with a novel scheduler called \DataflowNamenospace, which together enables identifying and exploiting reuse over complex DAGs of tensor operations. \AccelName provides 4x geomean speedup and 4x energy efficiency over state-of-the-art accelerators
across HPC workloads.
\end{abstract}


\section{Introduction}
\label{sec:introduction}

Recently, works like the Tensor Algebra COmplier (TACO) \cite{XXXTaco} have spurred research interest in generalized tensor applications (used in high-performance computing (HPC)), and their acceleration using Deep Neural Network (DNN) Accelerators~\cite{extensor,asgari2020alrescha,fdmax,cerebras,plasticine,eyeriss2016isca,kwon2018maeri,tpu-isca}. However, this generalization presents new challenges. Tensor-algebra applications often have much more complex dependency graphs between operations.
For example, \autoref{fig:dfg} shows the complex cascade (aka DAG) of tensor operations found in the conjugate gradient (CG) over two iterations of CG loop, which is a high value HPC solver rich in these complexities and a key workload of interest in this work.
Many such HPC applications have memory-bounded \textit{skewed GEMMs}.

Tensor operations (e.g., GEMMs) used within DNNs offered good \emph{reuse} opportunities because of their relatively cubic aspect ratios and large dimensions. Earlier DNN accelerators exploited this \emph{reuse} by searching the optimal schedule for each tensor operation independently to achieve maximum reuse for the networks they targeted at design time. 
Simultaneously, these schedules could be efficiently implemented using \emph{scratchpad} buffers that were explicitly controlled by the schedule to stage the intermediate \emph{tiles} of the data according to the traversal order within each GEMM\footnote{For explicitly managed buffers (e.g. scratchpads), buffer allocation aka tiling is an additional knob in scheduling search space.}.
There are two forms of data reuse in such a DAG that any scheduler could leverage: (i) within individual tensor operators and (ii) over the edges of the DAG. Earlier DNN accelerators exploited (i) by optimizing the schedule for tensor independently

Unfortunately, reuse within individual tensor operators is fairly limited in many HPC tensor operators, including CG, due to highly skewed shapes of the operators (\autoref{sec:apps}). This is due to an underappreciated yet fundamental property, namely, \textit{not all dense GEMMs with a large number of multiplications are compute bound even in the best case}. SkewedGEMMs are memory bound and leave datapath resources idle. 

\insertFigureScaled{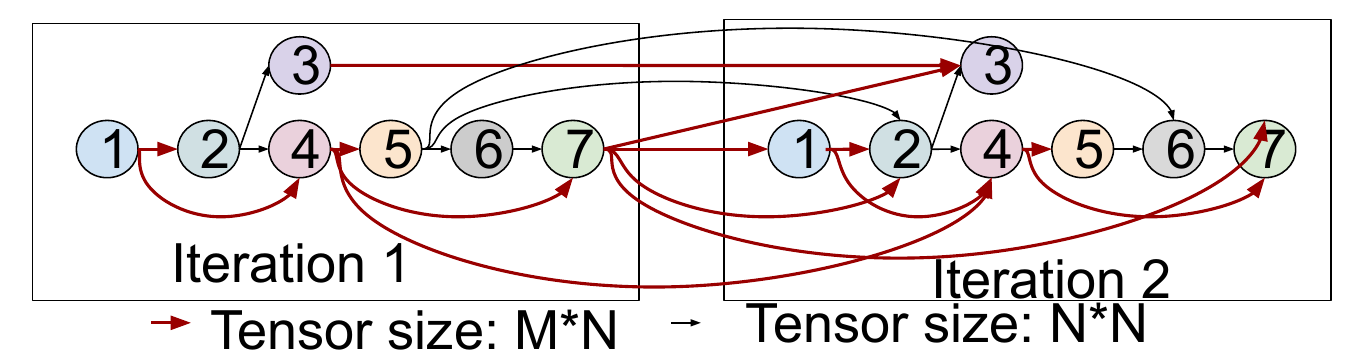}{Tensor dependency graph of intermediates in Conjigate Gradient across first two iterations of the CG loop.}{0.9}

Therefore, this work focuses on extracting reuse over the DAG edges (aka inter-operation reuse). 
Some recent works have demonstrated that \textit{pipeline reuse} between adjacent operators~\cite{tileflow,isca-pip,yan2020hygcn,garg2021understanding} can be leveraged by using on-chip \emph{scratchpads} to additionally stage the intermediate tensor, thereby reducing traffic to main memory.
Unfortunately, it is challenging to generalize inter-operation pipelining to complex DAGs since the transitive edges in complex DAGs introduce additional data dependencies, multicasts, diverse reuse distances across operators, and possible data layout transformations (\textit{aka} swizzle) depending on the consumer. 
Thus, extending this to arbitrary DAGs remains an open problem to the best of our knowledge. 
Scheduling linear DNN DAGs on the accelerator scratchpads is already known to be a highly challenging compiler problem due to the large scheduling space and layout choices~\cite{dnnfusion,TelaMalloc}.
At the DAG level, these compilers 
only perform fusion of tensor operators with adjacent element-wise operations~\cite{dnnfusion}, or perform tensor-tensor fusion (e.g., FlashAttention~\cite{FlashAttention}, FLAT~\cite{flat} and OMEGA~\cite{garg2021understanding}). 

The aforementioned DAG complexity explodes the scheduling space for finding good scratchpad configurations further, as the total combinations and proportions of allocations burgeon with operation DAG depth and the number of tensors involved. For example, for a 7-operator DAG in ~\autoref{fig:dfg} (one iteration), the number of buffer allocation choices alone goes from 7$\times 10^{15}$ in a baseline scenario with reuse only in individual tensor operations to $10^{80}$ when additionally considering reuse along edges of the DAG (\autoref{sec:arguments}). To make matters worse, scientific applications like CG have different tensor shapes for each problem, unlike DNNs, which means that costly buffer allocation needs to be done for every problem.
Hence, state-of-the-art accelerators that use explicit data orchestration cannot reuse along delayed downstream dependencies.

To this end, this work proposes an accelerator~\AccelNamenospace\footnote{\AccelNameexp}. \AccelName makes the scheduling problem tractable by introducing a novel explicit-implicit on-chip buffering mechanism called \SpadNamenospace\footnote{\SpadNameexp}, which we classify as a hybrid between explicitly-programmed scratchpads and implicitly managed caches, leveraging the strengths of both.
Specifically, \SpadName leverages high-level information (such as reuse distance and frequency, start and end addresses, indices) about the tensor operands from a software scheduler, but manages tensor placement and replacement via implicit (i.e., hardware-managed) policies. 
\AccelName uses our proposed scheduler, called \DataflowNamenospace\footnote{\DataflowNameexp}, which ``identifies" diverse reuse opportunities of the various operands and edges in an arbitrary DAG, which can then be leveraged by \SpadNamenospace. 
\DataflowName and \SpadName together enable~\AccelName to manage coarse-grained buffer management decisions at the tensor granularity (not cacheline) while enabling the hardware to manage track and leverage data reuse (removing this complexity from the compiler).

\DataflowName is more comprehensive than prior work in scheduling along edges of an arbitrary DAG as it also identifies reuse opportunities for all kinds of delayed downstream consumers, and minimizes the need to transform data layout (aka swizzle). The buffer allocation aspect of scheduling for complex dependencies is done by a hybrid implicit/explicit manner by~\SpadNamenospace, as opposed to buffer mechanisms in accelerators that do explicit buffer allocation~\cite{buffets}. \SpadName makes the problem tractable by reducing the complexity of the buffer allocation step from $10^{80}$ to $10^2$ (\autoref{sec:arguments}), since it only leverages high level DAG connectivity information and makes cycle-level decisions implicitly. On the other hand, \SpadName also significantly reduces the metadata area overhead of cache as it needs one entry per tensor, and replacement policies consider the tensor as a whole as opposed to individual lines leading to better policies than caches.
\AccelName achieves 4x geomean speedup and 4x improvement in energy efficiency across a broad range of scientific workloads 
over the state-of-the-art.

\section{Background}
\label{sec:background}

\subsection{Intra-operation Scheduling}


The following equation describes GEMM in the einsum representation~\cite{XXXTaco}:
    $Z_{m,n}=\Sigma_{k}A_{m,k}*B_{k,n}$

In equation (1), $A$, $B$ and $Z$ are the \emph{tensors} and $m$,$n$ and $k$ are \emph{ranks}. $k$ is a \emph{contracted dimension}, while $m$ and $n$ are \emph{uncontracted dimensions}. For brevity the summation $\Sigma$ can be omitted as the contracted ranks do not appear in the output tensor. 
Operations can be straightforwardly scheduled using concrete loop nests, for example:
\begin{footnotesize}
  \begin{verbatim}
M1=M/M0
//schedule A                 //schedule B
1 for m1 in range(M1):       for m1 in range(M1):
2  for n in range(N):         for k in range(K):
3   pfor k in range(K):        pfor n in range(N):   
4    pfor m0 in range(M0):      pfor m0 in range(M0):
      m=m1*M0+m0                 m=m1*M0+m0
5     Z(m,n)+=A(m,k)*B(k,n)      Z(m,n)+=A(m,k)*B(k,n)
\end{verbatim}  
\end{footnotesize}

These schedules apply a variety of loop transformations~\cite{timeloop} such as loop ordering, varying parallelization strategies and tiling (the tile size in the example is M0). The actual efficiency (datapath utilization and energy-efficiency) of each schedule depends on whether it can find enough reuse to reach the compute-bound regime of a given architecture.

\subsection{Inter-operation Pipelining}
\label{sec:inter-op-pipelining}
Inter-operation pipelining is a form of scheduling adjacent dependent operators in workloads with chains of Einsums (e.g., DNNs) such a way that the intermediate tensor output is consumed on-chip without having to be written to and read from main memory. Inter-operation pipelining is currently an active area of research for memory-bounded workloads~\cite{fused,isosceles,tileflow,flat}. \autoref{fig:no-fusion}(a) shows pipelining between two operations.
Unfortunately, simple pipelining is insufficient to capture more complex DAG dependencies as we motivate in the next section, and extend the state-of-the-art by broadening the scope of inter-operation scheduling to the entire DAG over multiple iterations.

\begin{scriptsize}
    
\begin{table}[t!]
\begin{scriptsize}

\begin{center}

  \center
  \caption{Performance of CG compared to Linpack (HPL) on Top5 supercomputers. Adapted from HPCG~\cite{hpcg2021}.
} 
    \label{tables:hpcg}
  \begin{tabular}{|l|l|l|l|l|}
    \hline
    \textbf{Supercomputer} & \textbf{HPL} &\textbf{HPCG} & \textbf{HPCG flops} &\textbf{HPCG:}  \\
     & \textbf{Pflops/s} &\textbf{Pflops/s} & \textbf{as \% of HPL } &\textbf{\% of peak}  \\
    \hline
1. Frontier & 1206 & 14.05 & 1.16\% & 0.8\% \\ \hline
2. Aurora & 1012 & 5.61 & 0.55 & 0.3\\\hline
3. Eagle & 561.2 & \multicolumn{3}{l|}{\revision{Not available}} \\\hline
4. Fugaku & 442.01 & 16 & 3.62\% & 3\% \\ \hline
5. Lumi & 379.7 & 4.587 & 1.2\% & 0.87\% \\ \hline
  \end{tabular}
\end{center}
\end{scriptsize}

\end{table}
\end{scriptsize}

\section{Motivation and Overview}

\subsection{Problem Demonstration for Conjugate Gradient}
\label{sec:apps}

Tensor-algebra applications are chains of einsums where tensors produced by earlier equations are consumed by later ones. This results in a \emph{tensor dependency graph}, dictating the high-level production/consumption of data throughout the HPC region of code. As a running example of how this affects reuse and utilization, we use the HPCG benchmark~\cite{dongarra2015hpcg,hpcg2021}, which runs Conjugate Gradient (CG)---a widely used HPC solver application represented as a DAG of tensor operations.

As \autoref{tables:hpcg} shows, \textbf{\textit{CG achieves only 1-3\% of peak performance on top 7 supercomputers}}. 
A key reason is due to these implementations relying on intra-operation reuse, and the skewed nature of the individual GEMM aspect ratios means they cannot find sufficient reuse. In the rest of this section we explain this phenomenon in detail.

\insertWideFigureScaled{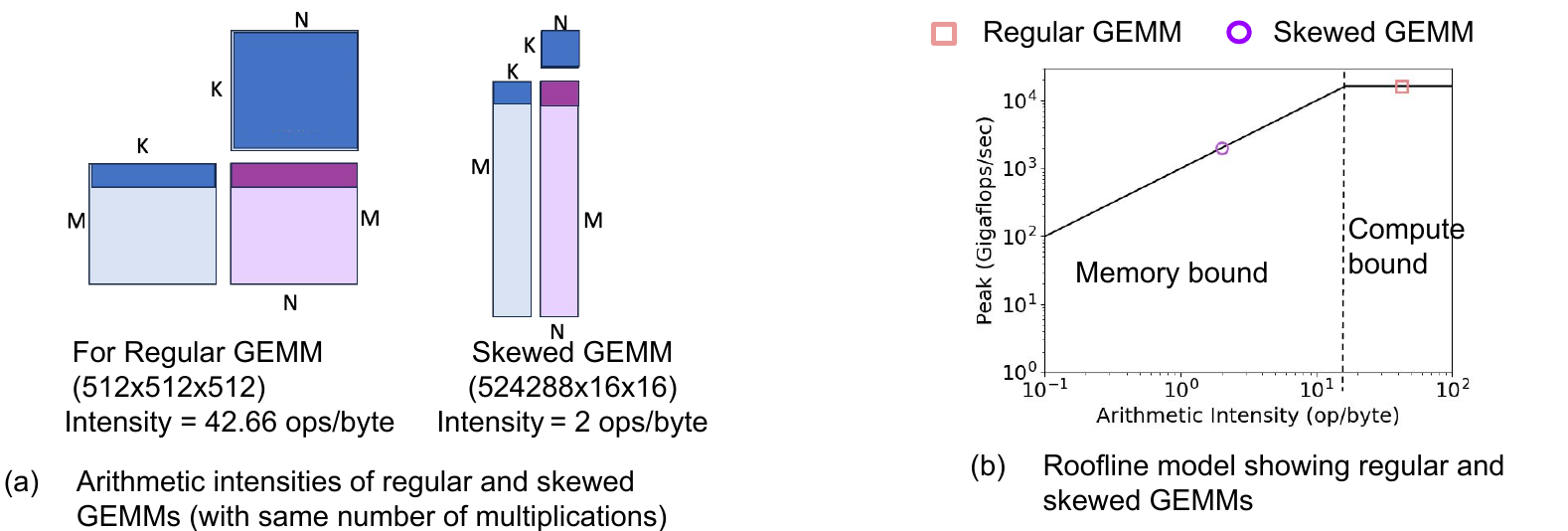}{Arithmetic intensity comparison and roofline plot for regular and skewed GEMMs. Word size = 32-bit and memory bandwidth  = 1TB/sec.}{0.8}

\indent \textbf{Sparse and Skewed GEMMs in Conjugate Gradient:~}
Iterative linear solvers solve the system of linear equations-

\vspace{-3.5mm}
\begin{equation}
    A_{m, k} * X_{k} = B_{m}
\end{equation}

While traditional CG considers b and x as vectors, block CG works on multiple initial guesses simultaneously for faster convergence, thus making it a matrix multiplication problem:

\vspace{-3.5mm}
\begin{equation}
    A_{m, k} * X_{k, n} = B_{m, n}
\end{equation}
\LinesNotNumbered

\begin{algorithm}
\begin{footnotesize}

\caption{Chain of operations for CG. $k$ represents contractions of size $M$ and $j$ represents contractions of size $N$ and $N'$. Line numbers represent significant computation steps referred to throughout the text.
}
\label{alg:cg_einsum}
\KwIn{$A_{m,m}$, $B_{m,n}$, $X_{m,n}$}
$R_{m,n} = B_{m,n} - A_{m,k} * X_{k,n}$ \\
$\Gamma_{n,n}=R_{k,n}*R_{k,n}$\\
$P_{m, n} = R_{m, n}$ \\
\While{not converged}
{
  \nl  $S_{m,n} = A_{m,k} * P_{k,n}$   \textcolor{gray}{~~~// SpMM.}  \\
 \nl   $\Delta_{n',n} = P_{k,n'} * S_{k,n}$ and $\Lambda_{n',n} = \Delta^{-1}_{n',j} * \Gamma_{j,n}$ \textcolor{gray}{~~~// $\Delta=P^TS$} \\
 \nl   $X_{m,n} = X_{m,n} $+$ P_{m,j} * \Lambda_{j,n}$ {~~~\color{Green}{// $X$ represents the initial guess which finally becomes the solution on convergence}} \\
\nl    $R_{m,n} = R_{m,n} - S_{m,j} * \Lambda_{j,n}$ {~~~\color{Green}{// $R$ represents the difference between B and AX, which in convergence should be sufficiently small}} \\
$\Gamma\_prev_{n,n}=\Gamma_{n,n}$\\
 \nl   $\Gamma_{n',n} = R_{k,n'} * R_{k,n}$ \textcolor{gray}{~~~~~~~~// $\Gamma=R^TR$} \\
\If{$all(diag(\Gamma)\leq \epsilon)$} 
{
    \textbf{break}
}
\nl    $\Phi_{n',n} = \Gamma\_{prev}^{-1}_{n',j} * \Gamma_{j,n}$ \\
\nl    $P_{m,n} = R_{m,n} $+$ P_{m,j} * \Phi_{j,n}$ {~~~\color{Green}{// $P$ represents the search direction}}
}
\textbf{return} $X$
\end{footnotesize}
\end{algorithm}
\vspace{-1mm}

\autoref{alg:cg_einsum} shows the tensor operations in the CG Algorithm. 

From the perspective of tensors, $P$, $R$, $S$ and $X$ (named using English-letter variables except A) are highly skewed ($M\times N$), for example $1000000\times 8$. In contrast, tensors like $\Delta$, $\Lambda$, $\Phi$ and $\Gamma$ (named using Greek letters) are small tensors ($N\times N'$), for example, of size $8\times8$ Also, $A$ is the only sparse tensor in CG with a maximum shape($M\times M$),  for example, $1000000\times1000000$ but with occupancy of 1-100 non-zeros per row. As a result, line 1 is a sparse SpMM operation while all the other matrix multiplication operations are dense w.


%

Notably, in CG applications, we observe that one dimension is too large and other dimensions are too small, thus leading to skewed aspect ratios. These skewed GEMMs have low data reuse. We quantify data reuse by using the metric \emph{arithmetic intensity}\cite{roofline} which is defined as number of operations per byte of data moved. Thus, low arithmetic intensity implies that the data moved could not be reused for enough operations.

In a GEMM where $M$ is the large dimension and $K=N$, the best possible arithmetic intensity can be calculated as follows:

\begin{equation}
  AI_{best}  = \frac{Number~of~MACs}{Minimum~DRAM~Accesses}
\end{equation}

For an individual operation with no inter-operation reuse, all tensor operands begin and end in DRAM and are accessed atleast once.

\begin{equation}
  AI_{best}  =\lim_{K/M\to0} \frac{M\times K\times N}{M\times K\text{ + }K\times N\text{ + }M\times N} = \frac{N}{2} ops/word
\end{equation}

For workloads where N <= 16 (e.g. CG), this translates to <=2 ops/bytes for 32B word size, and therefore degrades to memory-bound performance in isolation (see \autoref{fig:ai1}). This motivates the need to exploit inter-operation reuse to increase intensity in scientific applications.
\autoref{fig:ai1}(a) shows drastically lower arithmetic intensity for skewed GEMMs despite same number of multiplications. 

Roofline model (\autoref{fig:ai1}(b)) is used to plot the variation in throughput with arithmetic intensity. Based on this plot, performance can be memory-bound ie limited by memory bandwidth or compute-bound limited by datapath resources. As \autoref{fig:ai1}(b) shows, CG is highly memory bound even in the best case. Even though prior works on individual GEMM mapping~\cite{kwon2019understanding,timeloop} or CG accelerators~\cite{asgari2020alrescha} optimize individual operations, to achieve the best possible arithmetic intensity and achieve the roofline performance, the roofline performance itself is orders of magnitude less than the arithmetic intensity. 

Increasing "arithmetic intensity" ie reuse increases the performance until compute is fully utilized. Therefore, this motivates us to explore inter-operation reuse as intra-operation reuse alone is limited in case of CG.

\insertFigure{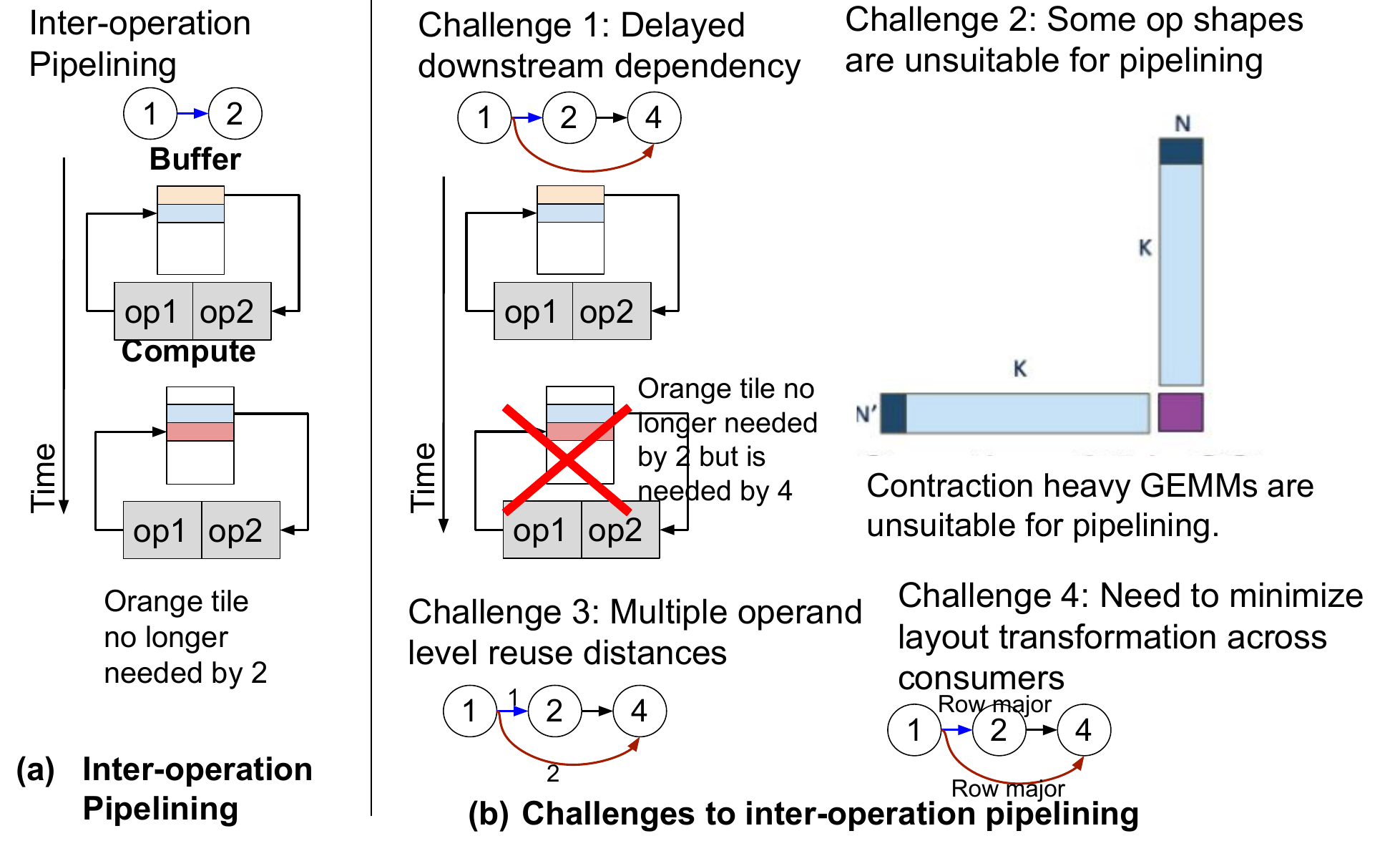}{(a) Inter-operation pipelining (b) Challenges to applying inter-operation pipelining to complex DAGs - 1) Delayed downstream dependency, 2) varying shapes, 3) consumers at multiple reuse distances 4) need to preserve layout across consumers.}

\subsection{Limitations and Challenges of Inter-operation Pipelining}
\label{sec:nofusion}




Effectively exploiting inter-operation pipelining (\autoref{sec:inter-op-pipelining}) 
has several challenges that we detail here.~\autoref{fig:no-fusion}(b) visually shows these challenges. 

\textbf{Challenge 1:} A delayed downstream consumer needs to use the intermediate operand. Traditional pipelining involves overwriting the previously produced tile when its no longer needed by the immediate consumer, which only works when there is no downstream consumer. Therefore these operands need to be stored in the on-chip or off-chip memory.

 \textbf{Challenge 2:} Not all layers pipeline with the next layer. For example, if a layer has large contraction (ie. lines 2 and 5 in~\autoref{alg:cg_einsum}), most of the compute is just spent on generating a usable output.

 \textbf{Challenge 3:} Once we have multiple downstream consumers, their on-chip reuse is a challenging problem given multiple reuse distances.

 \textbf{Challenge 4:} Given that the same operand has multiple consumers in a complex cascade, preserving the layout of the data in the on-chip memory is also crucial to exploit the reuse opportunity, and the scheduler needs to take that into account.

\begin{table*}[ht]
\begin{scriptsize}
    \centering
    \caption{\DataflowName comparison against prior schedulers, cost models and accelerators using their custom scheduling strategies.
    }
    \begin{tabular}{|p{0.29\linewidth}|p{0.03\linewidth}|
p{0.04\linewidth}|p{0.04\linewidth}|p{0.04\linewidth}|p{0.04\linewidth}|p{0.04\linewidth}|p{0.04\linewidth}|p{0.17\linewidth}|}
    \hline
       \textbf{Works} & \textbf{Intra-op reuse} & \textbf{Parallel Multicast} & 
       \textbf{Inter-op pipelining} & \textbf{Delayed hold dependency} & \textbf{Delayed writeback dependency} & \textbf{Swizzle Minimization} & \textbf{Part Implicit buffer} & \textbf{Scope of reuse given the dependencies} \\\hline
       MAESTRO~\cite{kwon2019understanding}, Timeloop~\cite{timeloop}, CoSA~\cite{cosa}, GAMMA~\cite{kao2020gamma}, Intersteller~\cite{interstellar}, TPUv1~\cite{tpu-isca}, MAERI~\cite{kwon2018maeri}, Eyeriss~\cite{eyeriss2016isca}, Eyeriss v2~\cite{eyeriss2}  & \greencheck &  \redcheck &
     \redcheck & \redcheck & \redcheck & \redcheck & \redcheck & Just within op-reuse. \\\hline
        FusedCNN~\cite{fused}, FLAT~\cite{flat}, FlashAttention~\cite{FlashAttention}, X-layer~\cite{xlayer}, ISOSceles~\cite{isosceles}, TileFlow~\cite{tileflow}, OMEGA~\cite{garg2021understanding} & \greencheck &  \redcheck & 
        \greencheck & \redcheck & \redcheck & \redcheck & \redcheck & Can reuse among adjacent ops only when there is no delayed dependency \\\hline
        SET~\cite{isca-pip}, TANGRAM~\cite{tangram} & \greencheck &  \greencheck & 
        \greencheck & \greencheck & \redcheck & \redcheck & \redcheck & Can reuse adjacent ops with delayed hold dependency\\\hline
       \DataflowName (This work)  & \greencheck  & \greencheck & \greencheck & 
       \greencheck & \greencheck & \greencheck & \greencheck & Can reuse adjacent ops with both delayed hold and writeback dependencies\\\hline
       
    \end{tabular}
    
    \label{tables:related_score}
\end{scriptsize}
\end{table*}

\section{\AccelName Accelerator Overview}
\label{sec:interface}


\autoref{fig:accelerator-figure} shows an overview of the overview of \AccelNamenospace. 
\AccelName leverages a configurable PE array that enables both temporal and spatial scheduling of multiple operators (as shown in \autoref{alg:cg_einsum}) inspired by prior works~\cite{planaria,sara_dac2022,plasticine,kwon2018maeri}.
Our key novelty is in the on-chip memory hierarchy to extract inter-operation reuse in arbitrary DAG of operations. The on-chip memory hierarchy consists of different buffers.  
A scratchpad (explicit) called \textit{input buffer} fetches input tensors from DRAM. An explicit \textit{pipeline buffer} is used to stage producers and consumer for operations where pipelining is possible (without DRAM being involved).
\SpadName is our proposed hybrid explicit-implicit buffer mechanism to reuse tensor operands that have to be written back where pipelining them is not possible.

We demonstrate how accelerator~\AccelName uses co-design of the \DataflowName scheduler (\autoref{sec:score}) and the~\SpadName buffering mechanism (\autoref{sec:chord}) to identify delayed downstream dependencies and exploit reuse opportunities among them in complex DAGs, without exploding the scheduling search space.~\autoref{sec:arguments} also discusses the shortcomings of caches and scratchpads which~\SpadName resolves.
Table \ref{tables:related_score} shows a detailed breakdown of existing state-of-the art scheduling approaches versus \DataflowName and how they respond to these challenges. 
Table \ref{tables:related_chord} shows a similar breakdown of existing and the \SpadName hardware buffering mechanisms, and how their approach is exposed to the scheduler or programmer.

\insertFigureScaled{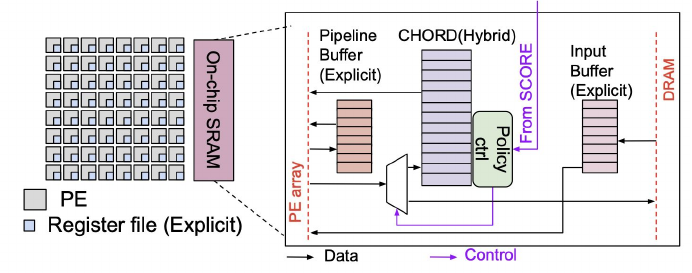}{
Overview of \AccelNamenospace. Within the memory hierarchy, input buffer, pipeline buffer and register files are explicitly managed while~\SpadName is hybrid implict-explicit. We expand on~\SpadName in~\autoref{sec:chord} (\autoref{fig:chord-fig}).}{1}

\section{Software Scheduler: \DataflowNameTitle}
\label{sec:score}

The~\DataflowName scheduler takes in the application represented as a DAG of tensor operations, as the input as~\autoref{fig:system} shows. It identifies types of dependencies at the tensor level and then determines the schedule of the operands that maximizes exploiting pipelining wherever it can and steers the operands with downstream consumers requiring writeback to~\SpadNamenospace. \autoref{tables:related_score} shows the scope of dependencies~\DataflowName can cover compared to previous schedulers.

\vspace{-1mm}
\begin{algorithm}[ht!]
\definecolor{LimeGreen}{HTML}{32CD32}
\definecolor{BrickRed}{HTML}{AA4A44}
\SetInd{0em}{1em}
\caption{Determination of tensor-level dependencies in a DAG}\label{alg:inter-op}
\begin{footnotesize}

    \For{node $\in$ NODES}
{   
node.numcast=0	\\
\For{edge $\in$ edges(node)}
{
    flag=0\\
      \If{edge.isTransitive == false}{
    node.numcast++\\
    \If{node.numcast > 1}
    {
        \textcolor{LimeGreen}{node.parallel\_multicast=true}
    }

}\If{node.dominance$\neq$contracted$\And$edge.isTransitive==false$\And$node.pathnext.dominance$\neq$unshared}
   {
edge.dependency=\textcolor{blue}{pipelineable}
   }
\If{node.dominance==contracted | node.op$\neq$tensor\_mac}
   {
edge.dependency=sequential
   }
   \If{edge.dest.dominance $\notin$ edge.tensor.ranks}
   {
edge.dependency=sequential
   }
   \If{node.dominance$\neq$contracted$\And$edge.isTransitive==true$\And$ node.pathnext.dominance$\neq$unshared}
   {
\For{pathnode $\in$ longestpath(edge)}
{
\If{pathnode.dominance==contracted or pathnode.pathnext.dominance == unshared}
{edge.dependency=\textcolor{BrickRed}{delayed\_writeback} \\ 
flag=1 \\
break
}
}
\If{flag==0}
{edge.dependency=\textcolor{cyan}{delayed\_hold}
}
   }
}
}

\end{footnotesize}

\end{algorithm}

\subsection{Tensor-level Dependencies in the DAG}
\label{sec:patterns}

We taxonomize different tensor-level dependences as follows.

\textbf{Sequential Dependency} - The sequential dependency implies that the nodes connecting that edge are executed one by one, although it may be possible to reuse the data on-chip depending on the space in the buffer. There are generally no constraints with this type of dependency and this is used when the dependency is not pipelineable. Operands are written to and later read from off-chip or on-chip memory. We call the dependency Sequential when the source operation does not pipeline with the adjacent destination operation. 



\textbf{{Pipelineable Dependency}} - A piplineable dependency refers to the edge of the DAG where pipelining used between the two operations. Prior works~\cite{flat,garg2021understanding,tangram,yan2020hygcn} make use of this. ~\autoref{fig:system} shows how pipelining reuses the data within SRAM in blue arrows. It discards the tile once it's consumed.
Pipelining does not provide benefit when the contracted rank is dominant (ie lines 2 and 5 in~\autoref{alg:cg_einsum}) because the bulk of compute is just used in producing data and that stage duration would become a rate limiting step.
 Pipelining completely removes the need to write and read the intermediate tensor. Most of the prior works including TileFlow~\cite{tileflow}, SET~\cite{isca-pip} exploit this reuse opportunity to pipeline the tensors wherever possible. However, there are additional delayed writeback or hold dependencies where there is a downstream consumer.

\insertFigureScaled{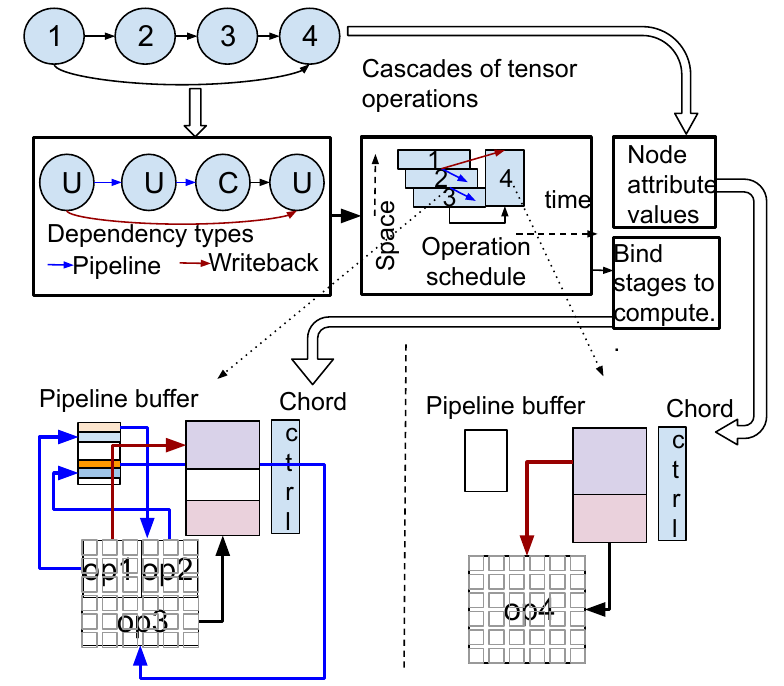}{Overview of~\DataflowName showing dependency classification, scheduling and binding attributes to~\SpadNamenospace. We show how the example schedule maps to hardware. 'U' and 'C' represent uncontracted and contracted ranks.}{0.8}

\insertFigure{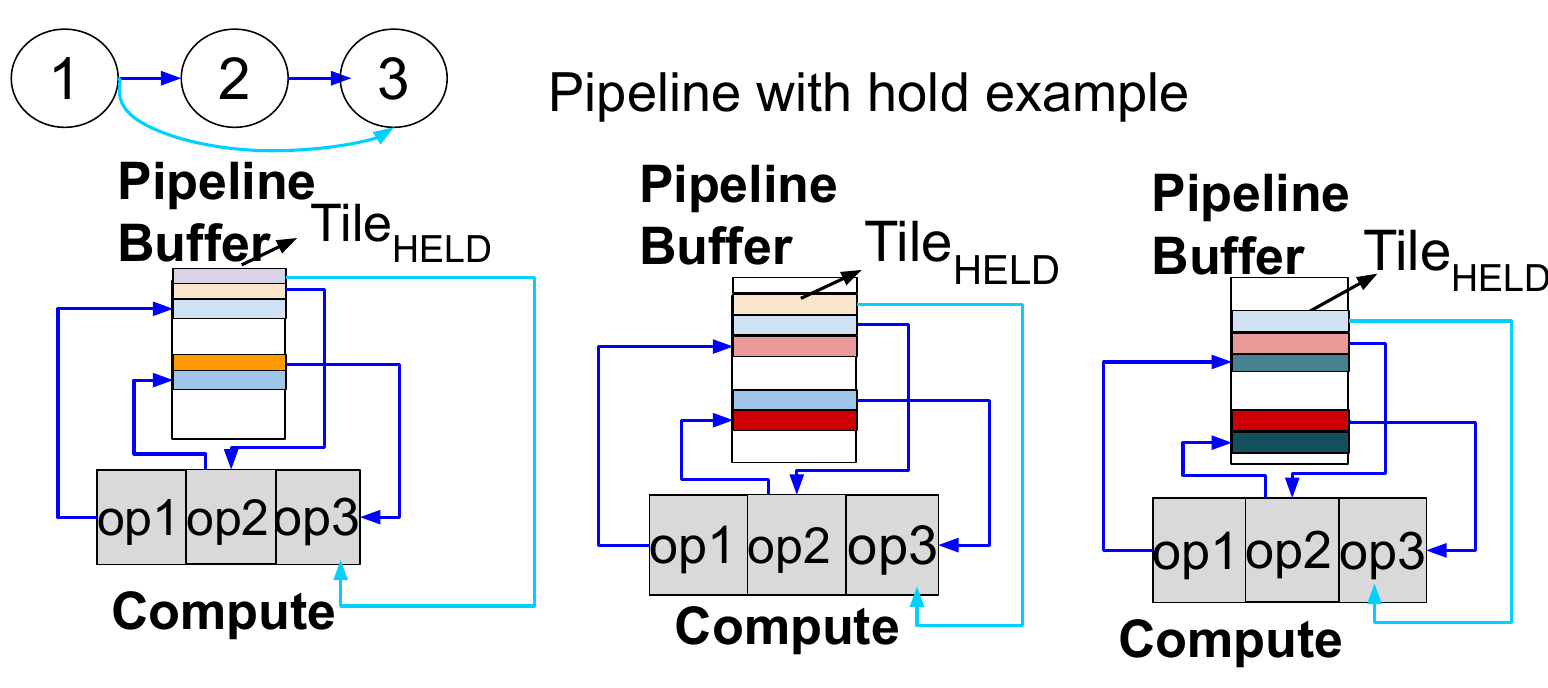}{Delayed\_hold buffer example.}

\insertFigure{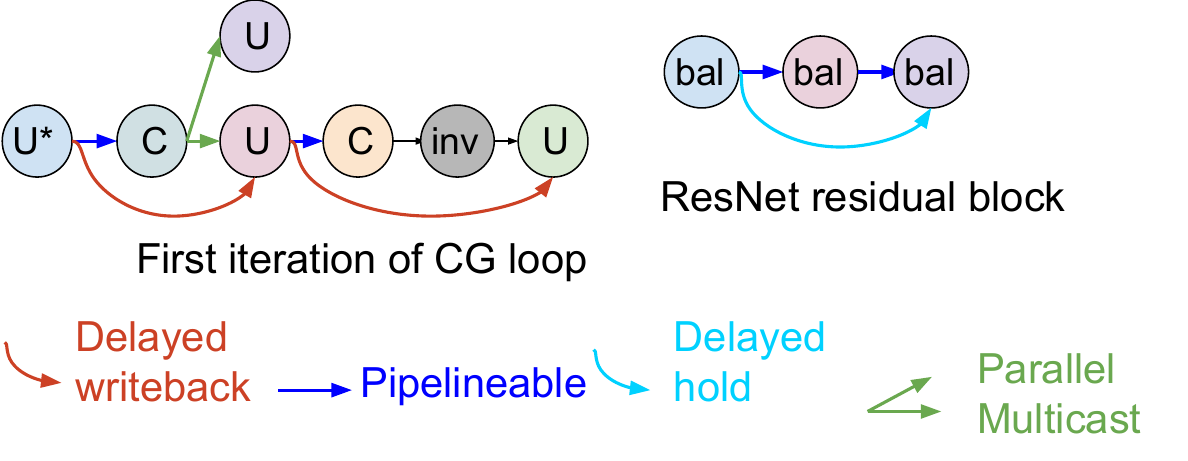}{Output of~\autoref{alg:inter-op} using colored edges. The letters in the node denote dominance, 'U' means uncontracted, 'C' means contracted, and 'bal' means that all ranks are big. The first operation is 'U' because the contracted rank is compressed.}

\textbf{{Delayed\_Writeback dependency}} -  In some cases, the intermediate tensor might be needed in a future computation, a dependency that has not been observed in previously popular applications like DNNs, GNNs etc. This is represented by the transitive edge on the graph. Therefore, as~\autoref{fig:system} shows in brick red arrow, the produced tensor needs to be stored in an on-chip SRAM buffer or DRAM depending on the capacity and tiles cannot be overwritten unlike simple pipelineable dependency. 
{Delayed\_writeback} dependency can be seen in CG as~\autoref{fig:inter-op-reuse} shows in brick red color.

\textbf{{Delayed\_Hold Dependency}} - In some cases with downstream consumers, there is a possibility that the complete chain until the future destination of this tensor is pipelineable. In that case, we hold the tile in the SRAM until the destination consumes that tile.~\autoref{fig:hold} shows an example of this. The number of tiles held essentially depends on the reuse distance of the downstream dependency (in terms of the number of operations). This dependency does not preclude us from reaping the full benefit of pipelining but requires slightly more occupancy in the on-chip SRAM. This dependency can be found in the ResNet~\cite{resnet} block with skip connections, as~\autoref{fig:inter-op-reuse} shows in cyan color. This is exploited by TANGRAM~\cite{tangram} and SET~\cite{isca-pip}.


\textbf{{Parallel Multicast}} - It is also possible to reuse the tensor in multiple parallel tensor operations, which we call {Parallel\_multicast}. These parallel operations are non-transitive edges from the multicasting nodes to the directly connected nodes.


\autoref{alg:inter-op} shows the methodology to mark the classify the dependencies for arbitrarily connected DAG of tensor operations. 
A non-transitive edge\footnote{A transitive edge is the edge not on the longest path between the source and the destination} where the source node is uncontracted dominant and the destination is unshared dominant has a~{pipelineable} dependency. Note that~{pipelineable} dependency does not guarantee pipelining since it also depends on the loop order of the individual operations as we discuss in \autoref{sec:loop}. Any edge where the source node is contracted dominant or the destination is unshared dominant, has a sequential dependency. A transitive edge with uncontracted dominant source node could have~{Delayed\_hold} dependency, if all the edges on the corresponding critical path are pipelienable, otherwise it has the~{Delayed\_writeback} dependency. ~\autoref{fig:system} (first box within~\DataflowNamenospace) and
~\autoref{fig:inter-op-reuse} show example outputs of~\autoref{alg:inter-op}.

\insertFigureScaled{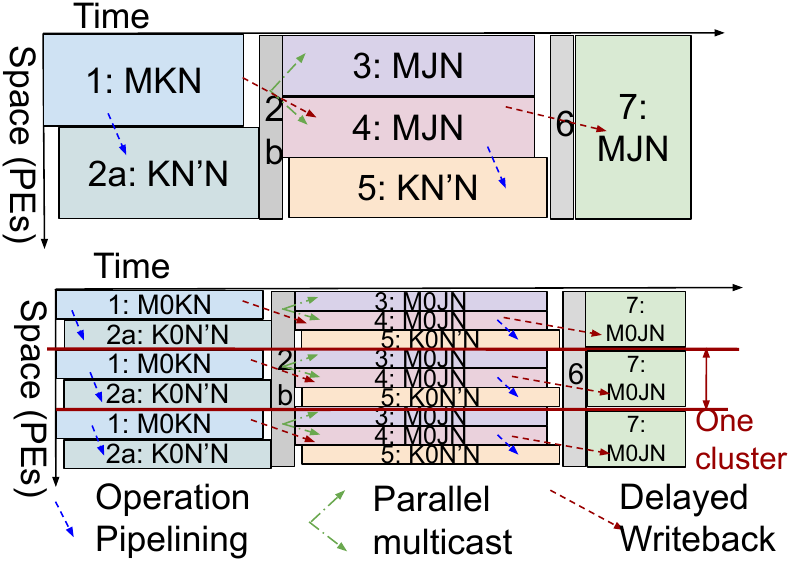}{CG iteration schedule. On top, the overall compute is divided into multiple parts each executing one tensor. At bottom, the tensors are split across the node by the dominant rank and the sub-tensors are pipelined within a cluster. Refer to~\autoref{alg:cg_einsum} for rank names and line numbers.}{0.8}

\subsection{Scheduling Operations}

\label{sec:loop}

\DataflowName schedules the operands, based on the dependencies and the DAG, such that pipelining is possible for maximum pipelineable edges. ~\autoref{fig:system} (second box within~\DataflowNamenospace) and~\autoref{fig:spacetime} show examples of the scheduler output.

\textbf {Loop orders}: 
If the edge has a pipelineable or Delayed\_hold dependency, the schedule tries to satisfy the codependence conditions of producer and the consumers in pipelining. 

For an example pair of Einsums $Z_{m,n}=A_{m,k}*B_{k,n}$ and ${W_{o,n}=C_{o,j}*Z_{j,n}}$, the conditions for pipelining a loop order pair are as follows-
\squishlist
\item The edge has a {pipelineable} inter-operation pattern.
\item Source node has uncontracted rank as the outermost loop.
\item Destination node has shared rank as the outermost loop. 
\item The shared tensor $Z$ is not swizzled since the same portion of data that is produced, should be consumed. 
\squishend

If the edge has a Delayed\_writeback or sequential dependency, the schedule tries to minimize layout transformation (swizzle) of a tensor, among various consumers.

We also keep the dominant rank in the outermost loop. Hence, large tensor is stationary and the small tensor is streamed from the register file.

\vspace{-1mm}
\begin{table*}[ht]
\begin{scriptsize}
    \centering
    \caption{\SpadName comparison against known buffer mechanisms. Most undesirable and generally undesirable properties are shaded red. Orange represents somewhat undesirable properties.}
    \begin{tabular}{|p{0.08\linewidth}|p{0.08\linewidth}|p{0.08\linewidth}|p{0.1\linewidth}|p{0.04\linewidth}|p{0.05\linewidth}|p{0.05\linewidth}|p{0.34\linewidth}|}
        \hline

       \textbf{On-chip mechanism}  & \textbf{Architectural Exposure} & \textbf{Placement Granularity} & \textbf{Placement Policy wrt Workload} & \textbf{Online Policy} & \textbf{HW Overhead} & \textbf{SW Burden} &  \textbf{Remarks} \\\hline

      Cache   & Implicit & \cellcolor{red!16}Line-level & \cellcolor{red!16}Fully Agnostic & \greencheck &  \cellcolor{red!16}Highest &   Lowest & Cache policy decisions are workload agnostic, with myopic line-level replacement and high fine-grained tag overheads. \\\hline


      Scratchpad   & Explicit  & \cellcolor{red!16}Line-level & \cellcolor{red!16}Fully Controlled and no support for dependencies &  \cellcolor{red!16}\redcheck & Lowest & \cellcolor{red!16}Highest & Data in the local address map needs to be fully controlled by the programmer, requires offline programming. \\\hline

       Buffets~\cite{buffets}  & Explicit & Tile-level (credit-based) & \cellcolor{orange!20}Fully Controlled &   \cellcolor{red!16}\redcheck  & Low &  \cellcolor{red!16}High & Buffets uses credit management scoreboarding to relatively ease synchronization over scratchpads.  \\\hline

       Tailors~\cite{tailors}  & Hybrid & Tile-level, Word-level & \cellcolor{orange!20}Fully controlled except overbooked tiles & \greencheck & Low& \cellcolor{red!16}High & Tailors uses buffets, which is explicit, except for managing overbooking, where the last entry (word-level) is implicitly managed and replaced.\\\hline

       
       \SpadName(This work)  & Hybrid (Coarse-grained explicit, cycle-level implicit) & Object-level & Object-aware policies with non-burdensome coarse-grained control &  \greencheck & Low & Low & \SpadName has cycle-level implicit replacement, incurs low area overhead, with low burden on programmer. Only information needed from workload (explicit) is the start and end addresses of tensors in global address map and the local address maps along with DAG information.  \\\hline
    \end{tabular}
    
    \label{tables:related_chord}
    \end{scriptsize}
\end{table*}

\textbf{Tiling:} 
Skewed GEMMs consist of two small dimensions, and as a result, have one small tensor. The small tensors are stored inside the registered file (explicit), and streamed from there. On the other hand, the large tensors are stationary. As a result, we keep the large dimension in the outermost loop. Even though, the register files are explicit, they do not require scheduling search, since we fix the mapping to stream the small tensor from the register file (where it fits entirely). This schedule already achieves the best-case intra-operation reuse and the best possible inter-operation pipelining.

However, for delayed downstream consumers, \DataflowName by itself does not do a fine-grained buffer allocation, as there are multiple downstream tensors contending for the space in~\SpadNamenospace. \DataflowName just provides~\SpadNamenospace's policies coarse-grained metadata about each operand, and the~\SpadName policies make fine-grained buffer allocation decisions.~\autoref{sec:arguments} explains the cost of search for the delayed downstream consumers.



\textbf{Handling sparsity} We store the sparse tensor in compressed (CSR/CSC) format, and we tile based on occupancy. Our tiling achieves the best possible arithmetic intensity for the individual SpMM operation.~\SpadName stores the data and the metadata in CSR/CSC format.


\textbf{Scalable Dataflow:} To ensure scalability for multiple nodes, the schedule also makes sure that pipelining is done within a node and not split across nodes in a multi-node architecture. In case of multiple node, we parallelize the dominant rank across the nodes. ~\autoref{fig:spacetime} bottom row shows the scalably tiled schedule. This makes sure that we are moving the small tensors across nodes instead of the large ones, as multiple tiles of large dimension will share the same small tensor. For example, if pipelining between operations 4 and 5, the top strategy in~\autoref{fig:spacetime} will require moving $SIZE_R$ data ie $M\times N$ words through the NoC. The bottom stragegy will require moving ($SIZE_{\Lambda}\times HOPS_{broadcast}$)+($SIZE_{\Gamma}\times HOPS_{Reduce}$) ie $N\times N\times $($Hops_{Broadcast} $+ $Hops_{Reduce}$). 

$M>>N$ in CG, and $M>>hops$ since one core can hold large M size, so the number of cores on which the execution is distributed is obviously way lower than $M$.


\subsection{\DataflowNamenospace-\SpadName Interface}

\DataflowName determines the type of buffer used in the memory hierarchy based on the dependency.
 We use the \textit{pipeline buffers} for delayed\_hold and pipelineable dependencies (\autoref{fig:system} and~\autoref{fig:hold}), by allocating slightly more space. 

The rest of the operands (sequential and writeback dependencies) are completely written back and required entirely in a downstream consumer therefore use~\SpadNamenospace (\autoref{fig:system}).~\autoref{sec:chord} describes the architecture of~\SpadName for maximally reusing such operands. Exploiting reuse in cases of pipelineable and Delayed\_hold dependencies also enables us to reduce the number of contending tensors for~\SpadNamenospace.


\insertFigure{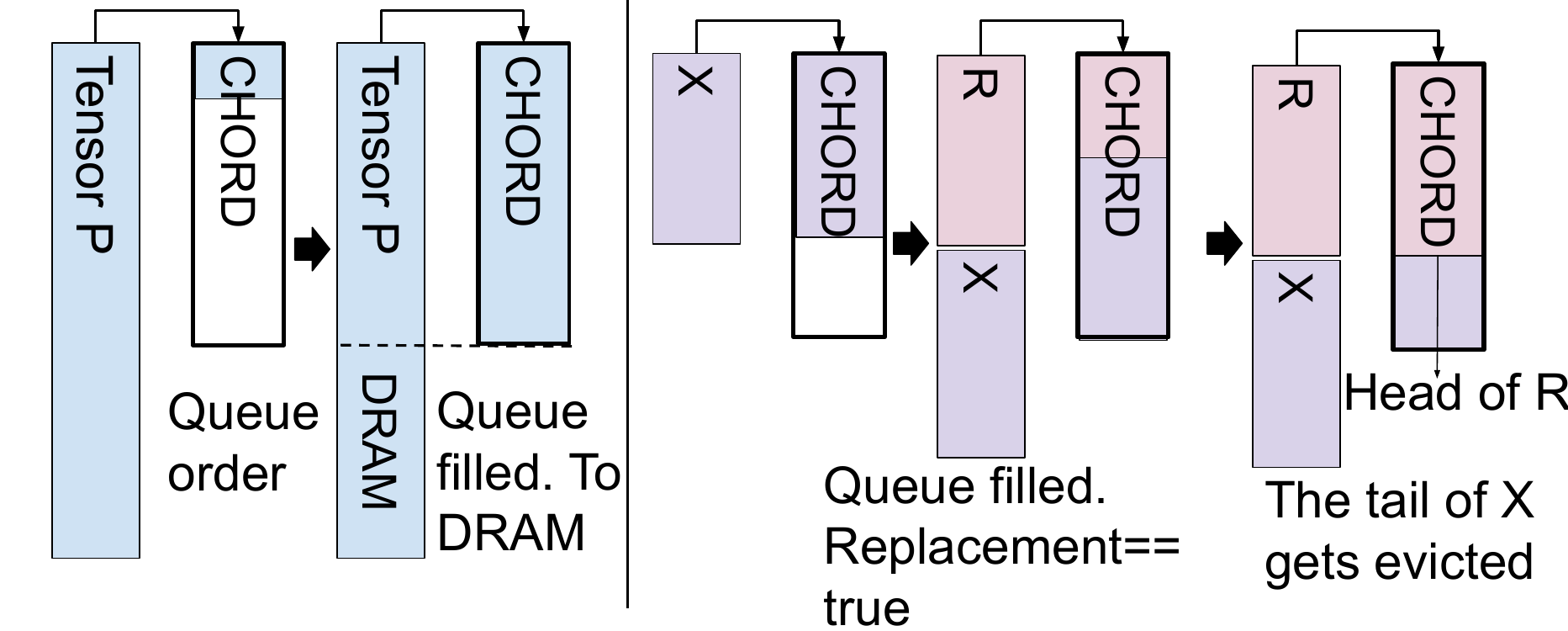}{~\SpadName policies. Left half shows~\PolicyA which starts by filling up without replacement. The part that could not fit is sent to DRAM. Right half shows~\PolicyB. After queue is filled, the tail of $X$ is evicted and next element of $R$ is enqueued at the head to fill the space.}

\insertWideFigureScaled{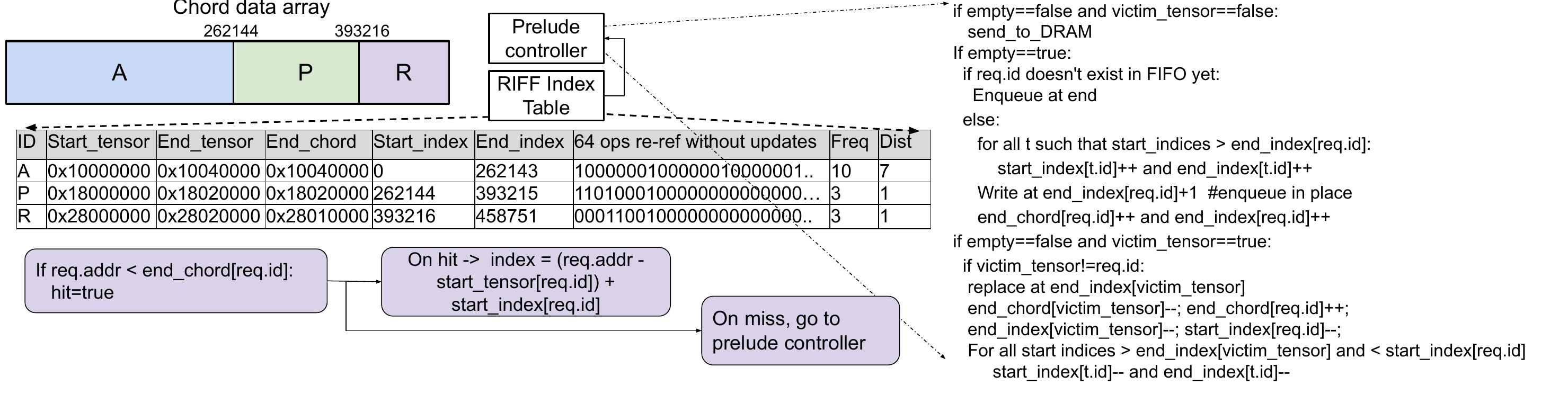}{Hardware mechanism \SpadNamenospace.}{0.95}

\section{Hardware Buffer Mechanism: \SpadNameTitle}
\label{sec:chord}

We propose~\SpadName for getting reuse out of tensors within the SRAM in a cycle-level implicit manner with coarse-grained explicit information from~\DataflowNamenospace.
In~\SpadNamenospace, we propose two replacement policies, \PolicyA and~\PolicyBnospace, which replace at an operand level rather than a cache line level. This eliminates the overhead of maintaining counters 
and tag matching for each line.  We also highlight the challenges with existing cache replacement policies and explicit scratchpad management. \autoref{tables:related_chord} summarizes \SpadName properties compared to prior buffering mechanisms.

\subsection{\SpadNameTitle~ Policies and Hardware Mechanism}


\label{sec:tornado}
We discuss ~\SpadName's two policies here.

\textbf{\PolicyA}- 
We write the tensor in the queue order and if the scratchpad is full, we send the remaining portion of that tensor directly to DRAM as~\autoref{fig:tornado} shows. This makes sure that we access the portion of the tensor in the scratchpad first rather than replacing it by the later part of the tensor and again reading the earlier part. This is in stark contrast to the cache replacement schemes that keep the recently accessed data. If the tensor fits, we repeat this for multiple operations. We read the tensor at head.

~\textbf{\PolicyB}- While~\PolicyA makes sure that we minimize the off-chip traffic for that particular tensor, there is still a chance that a particular tensor which is not reused until later (for example, $X$ in line 3 of~\autoref{alg:cg_einsum} which is used in line 3 of the next iteration) is written to the buffer at the expense of a future tensor that is re-used more frequently and before the reuse of the previous tensor (for example $R$ in line 4 which is reused in lines 5 and 7 of the same iteration). Based on the reuse frequency, we prioritize $R$ and evict out a portion of $X$ to make room for $R$ by evicting the later elements of $X$ first. We call this policy~\PolicyB. Then $R$ is read at its head. This is the first work to the best of our knowledge that considers reuse distance and frequency of a tensor in the dependency graph and proposes replacement at an operand granularity rather than line granularity. 




\autoref{fig:chord-fig}
shows the HW organization of~\SpadNamenospace. The buffer mechanism consists of a~\PolicyA controller which is responsible for detecting spills and send the spilling elements from the same operand to DRAM. While considering a different operand, \PolicyB index table determines whether the tensor must be replaced or not. Note that this index table will have far fewer entries than the total number of lines with a counter for each line. This table also keeps track of the starting and ending addresses of the slice of tensor residing in the SRAM and the ending address of the complete tensor.
The tensors in~\SpadName are contiguous and ordered. We also keep track of starting and ending index of the tensor as a part of per-tensor metadata. Thus, on hit, we can use this order and continuity to calculate the index in the buffer based on the request address without searching for it. \textit{This eliminates the need for tag matching per line.}
The information about the tensor's future reuse is obtained from~\DataflowName and is also part of the meta data used by~\PolicyBnospace. 
On a miss, we first check whether there is empty space by checking all end indices of existing tensors. If not, we check whether the tensor can replace any of the existing tensors, using the~\PolicyBnospace. If there is no replacement possible, we just send the data to DRAM. If there is a tensor element that can be replaced (which does not belong to the same tensor as the written tensor), we replace one entry from its tail. This is done by enqueuing the end index of the written tensor and letting the other elements of the written tensor push out the tail of the victim tensor.


\insertFigure{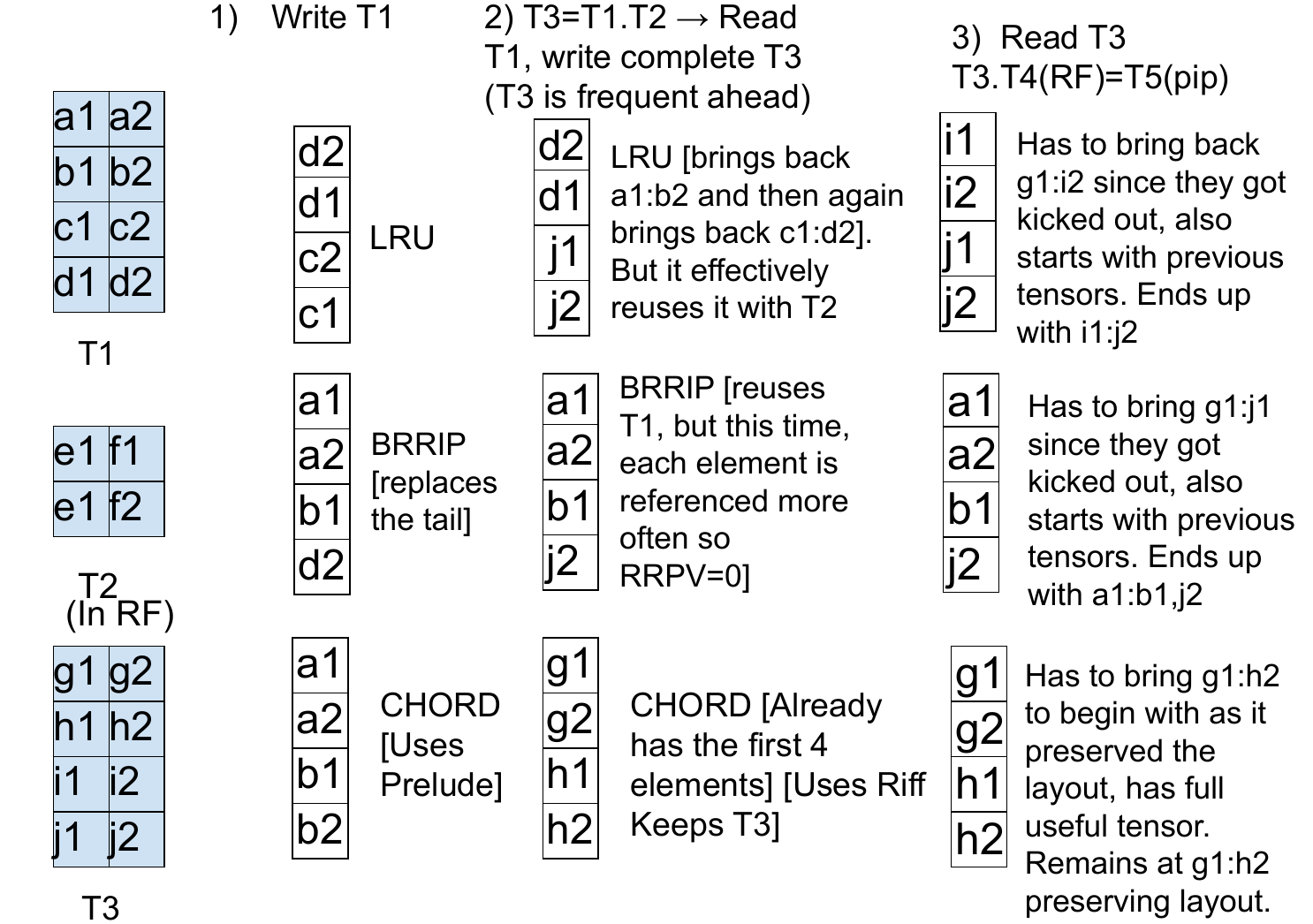}{Benefit of~\PolicyA and~\PolicyB over LRU and BRRIP.}

\subsection{Why \SpadName over Cache and Scratchpad?}
\label{sec:arguments}

\textbf{Better replacement than line-level cache replacement policies:}
Policy-wise, the main drawback of caches is the myopic view of lines which misses the tensor-level reuse opportunities.
\autoref{fig:replacement} shows a toy example to demonstrate the effectiveness of~\SpadName for tensor programs. In the first step, when a tensor is written to~\SpadName, it only keeps the head of the tensor (\PolicyAnospace) as that will be re-referenced first, while LRU kicks that out and then has to bring that in again. BRRIP~\cite{rrip} also keeps the first three data to prevent scanning behavior. In the next step,~\SpadName sees the reuse of the tensor T3 and overwrites it (\PolicyBnospace), while LRU and BRRIP have lingering elements from T1 as well, and the elements from T3 are the elements that will be re-referenced at the end. In the final step, LRU and BRRIP need to bring g1,g2... back, while~\SpadName already has them. Thus operand-level replacement is beneficial for such tensor programs.

\textbf{Hardware overhead reduction over caches:}
In terms of area, cache controller and tag bits are significant overheads and can account for almost a third of the cache area. Scratchpads on the other hand, do not incur control and tag area overhead, and the overall overhead is about 2\% compared to the data array size~\cite{buffets}. ~\SpadName needs only one entry of 512 bits per tensor (which is sufficient to include the metadata fields \autoref{fig:chord-fig} shows -- tensor ID, three addresses, two indices, tensor-level reuse history and reuse frequency and distance) to hold information about 64 tensors and their future operations in the DAG. This amounts to 0.01x of the area of the tag array inside the cache (4MB, 8-way associative). 

In terms of on-chip energy, tag access energy is comparable to data access energy, because of the size of the tag array and also due to set associativity. Whereas \SpadName completely minimizes this energy by taking advantage of contiguity in tensors and calculating the local address by just knowing the 512 bit metadata entry corresponding to the tensor in question, so that major energy comes from data accesses.

\textbf{Lower complexity of \SpadName compared to caches:}

\SpadName is less complex compared to the cache, \textit{in terms of both determining hit or miss, and changes in metadata.}

\textit{(1) Determining a hit or miss}: In a cache, determining hit or miss requires tag matching per line for every access. On the other hand, \SpadName uses bulk tracking at the tensor level where the tensors are contiguous, thus avoiding the need for tag matching per line. Hit or miss is determined by comparing the requested address with the end\_\SpadName address of requested tensor and that requires accessing a table that is 100x smaller than cache metadata.

\textit{(2) Changes to metadata:} In a cache, every hit or miss leads to a change in the recently used line, which requires updating the counters.
On the other hand,  \PolicyA and \PolicyB minimize the tracking required in \SpadName. On a hit, no change to the metadata is required, and the index calculation is performed based on start index and the start address of the tensor, which is obtained by reading from the \PolicyBnospace-index table, which is 100x smaller than cache metadata.
On a miss, the priority of the requested tensor is determined based on \PolicyBnospace, and the tail of the tensor tensor with the lowest priority is replaced. While this is happening, the start and end indices and end-\SpadName address associated with specific tensor IDs change, but that only corresponds to one entry per tensor in the \PolicyBnospace-index table. However, if the requested tensor has lower priority than all the other tensors in the \SpadNamenospace, the tensor is sent straight to DRAM, and there is no change within \SpadNamenospace.

Therefore, \SpadName avoids complexity by (1) avoiding tag matching per line, and instead only accessing "end-\SpadNamenospace" for the requested tensor from 100x smaller \PolicyBnospace-index table, and (2) changing start and end indices and end-\SpadName address in \PolicyBnospace-index table, only when there is a miss and the tensor is higher priority than atleast one tensor present inside \SpadNamenospace.

\textbf{High cost of scratchpad allocation solved by~\SpadNamenospace.}


The cost of programming a scratchpad for delayed operand reuse (with five tensors as an example) can be modeled in terms of number of buffer allocation choices as follows-


(1) The cost of dynamically allocating buffer to different slices of tensors.

$T1_{slice}+T2_{slice}+T3_{slice}+T4_{slice}+T5_{slice}<size$

$Num_{choices} = \frac{(size+4)!}{size!4!}\approx size^4$

For a 4MB buffer (assuming 32-bit data), this is approximately $\mathit{10^{80}}$ and it scales $\mathit{size^{T-1}}$ for $T$ tensors. Compared to this, the size of the map space is $10^{15}$ for op-by-op buffer allocation.

(2) Cost of arranging all the lines is $size!$ which must be multiplied. However, assuming that a tensor block is contiguous can bring this to $T!$, which is 120 for five tensors.

(3) Choosing the elements of the slice itself is expensive.

For each $Ti_{slice}$, there are $\frac{Ti!}{Ti_{slice}!(Ti-Ti_{slice})!}$ choices. This has a \textit{factorial space complexity} which is bigger than exponential complexity. However, assuming a contiguous block of elements chosen in a slice, this can be brought down to $Ti-Ti_{slice}$. For T tensors, this is raised to power T.

(4) Considering, the varying aspect of scratchpad allocation, a realistic constraint equation is 

${T1_{slice}(t)+T2_{slice}(t)+...+T5_{slice}(t)<size}$

In a reasonably modeled scratchpad, allocation of tensors changes with each access as we move through the program. Therefore the choices in (1),(2),(3) need to be raised to the power of number of time steps. Moreover, these costs are just for buffer allocation, excluding scheduling knobs, for example, loop order, parallelism etc.

Hybrid explicit-implicit data orchestration helps minimize the cost of static buffer allocation.
\PolicyB solves step (1) by dynamically controlling buffer allocation of the tensors.
\PolicyA co-designed with~\DataflowNamenospace's swizzle minimization solves step (3).~\SpadName as a whole being cycle-level implicit and coarse-grained explicit avoids the cost of (4). \revision{The design-space of~\SpadName corresponding to the buffer allocation step only corresponds to the~\PolicyB policy decisions based on tensor-level reuse distance and frequency is $O(nodes+edges)$ since it only needs high level DAG information, e.g. reuse distance. Normally, the number of nodes looked ahead by~\SpadName are of the order of $10^2$, which is drastically lower than $10^{80}$.}

\begin{scriptsize}
\begin{table*}[h]
    
\begin{scriptsize}
\begin{center}

  \center
  \caption{Different schedules and buffers configurations evaluated with the corresponding SOTA accelerator works.}
  \label{table:dataflow_config}
    
\begin{tabular}{|p{0.11\linewidth}|p{0.10\linewidth}|p{.15\linewidth}|p{0.53\linewidth}|}
    \hline
    \textbf{Schedule} & \textbf{Buffer-hierarchy} & \textbf{Combined Configuration} & \textbf{Description of the Combination of Schedule and the Buffer-hierarchy}  \\
    \hline
    Best Intra-layer   
 & Explicit  & Flexagon-like (Flexagon)~\cite{flexagon} & Flexagon's flexible architecture can optimally map a single (Op-by-op) operation with any shape and sparsity. This combination achieves  the best possible single operation reuse. All ops begin and end in DRAM. Its the oracle operation-by-operation dataflow.\\
    \hline
{Best Intra-layer} 
 & LRU Cache  & Flexagon with LRU (Flex+LRU) & All accesses go through the LRU cache without any explicit management.\\\hline

{Best Intra-layer}  
 & BRRIP Cache  &  Flexagon with BRRIP (Flex+BRRIP) & All accesses go through the BRRIP cache without any explicit management.\\\hline

    Pipelining  & {Explicit} & FLAT-like (FLAT)~\cite{flat} & We model the FLAT R-gran dataflow for pipelining between two operations when it is possible to apply (instances with delayed downstream consumers are not considered as pipeline just consumes the tensor without writeback). Also note, that this work assumes Parallel Pipeline (PP) throughout, and this baseline captures the dependency of intra-operation dataflows according to FLAT R-Gran, even though, the actual hardware implementation of FLAT uses Sequenatial Pipeline (SP) dataflow~\cite{garg2021understanding}. However, this does not impact the DRAM accesses. \\\hline

    

    \DataflowNamenospace & \SpadNamenospace & \AccelNamenospace\ &(\textbf{This work}) We use~\SpadName buffer for reuse of large tensors that have downstream consumers along with the explicit pipeline buffers,  and  RFs.  It uses both~\PolicyA and~\PolicyB policies.
    \\\hline
    \hline

    Best Intra-layer & \PolicyAnospace-only & \PolicyAnospace-only & \textbf{(Additional study in \autoref{sec:sens})} We turn off all other optimizations, and model the effect of an SRAM with \PolicyA as the only policy.\\\hline

     Pipelining + Delayed Hold & Explicit & SET-like (SET)~\cite{isca-pip} & \textbf{(Additional study in \autoref{sec:sens})} We add another baseline SET which supports the delayed hold, for the ResNet evaluation since that is the only place the dependency shows up. \\\hline
    
    
  \end{tabular}\end{center}



\end{scriptsize}

\end{table*}

\end{scriptsize}

\begin{scriptsize}
\begin{table}[t!]
\begin{scriptsize}

  \center
  \caption{Configurations accelerator architecture.
  } 
  \label{table:config}
  \begin{center}
  \begin{tabular}{|l|l|}
    \hline
   

    \textbf{Accelerator Architecture Parameters} & \textbf{Values}  \\
    \hline
    SRAM size & 4MB (\autoref{sec:sens} sweeps \{1,16\}MB)\\ \hline
   Number of MAC units & 16384\\ \hline

    Cache line size & 16 B\\\hline
    Cache Associativity & 8-way\\ \hline

    Memory bandwidth & 250,1000 GB/s \\\hline
    Clock Frequency & 1 GHz 
    \\\hline
       \SpadNamenospace's \PolicyBnospace-index table & 512 bits and 64 entries
       \\\hline

  \end{tabular}
\end{center}
\end{scriptsize}

\end{table}
\end{scriptsize}

    
\begin{table}[t]
\begin{scriptsize}
  \begin{center}
  \caption{Workloads and Datasets evaluated. For GCN, M refers to vertices and N and O refer to input and output features.
  }
  \label{table:dataset}
\begin{tabular}{llll}
\toprule
\textbf{Workload}   & \textbf{Dataset} & \textbf{Shapes and Sparsity} \\
\toprule
2D/3D problem & \reviewme{fv1} & \reviewme{M=9604, nnz=85264}\\\hline
Fluid Dynamics & \reviewme{shallow\textunderscore water1} & \reviewme{M=81920, nnz=327680}\\\hline
Circuit sim & \reviewme{G2\textunderscore circuit} & \reviewme{M=150102, nnz=726674}\\


\hline

GCN & cora & M=2708, nnz=9464, N=1433, O=7\\
Layer & protein & M=3786, nnz=14456, N=29, O=2
\\




\bottomrule
\toprule
ResNet (\autoref{sec:sens}) & ImageNet & Conv\_3x residual block 1~\cite{resnet}\\
\bottomrule
\end{tabular}
\end{center}
\end{scriptsize}
\end{table}

\begin{table}
\begin{scriptsize}
  \begin{center}
  \caption{Workloads parameters
  }
  \begin{tabular}{|l|l|}
    \hline
    \textbf{Workload Parameters} & \textbf{Values}  \\
    \hline
    Bytes per word/element in CG and GNN & 4B (32 bits)\\\hline
   \reviewme{Bytes per word in ResNet}   & \reviewme{2B (16 bits)}\\\hline 
   
     Iterations of CG loop & 10\\\hline
    N rank in CG & Sweep: 1, 16\\\hline

  \end{tabular}
\end{center}
\end{scriptsize}

\end{table}

\textbf{Scratchpads are not panacea for statically known DAGs:} {Statically known DAG does not automatically imply that the costly buffer allocation only needs to be done once.} 
One of the preconceptions related to the use of scratchpads is that, the programmer just needs to go through the costly buffer allocation only once, and it keeps serving the purpose, given that the DAG of tensor operations is known. This is true only if the workload is sufficiently compute bound, and the matrices are relatively cubic to allow the same mapping to be applied for all operations and across all problem sizes.

For memory bound tensors in CG that are competing for space in the buffer, the ideal proportion of tensor allocation does in fact change with the problem size. This means that for every new CG problem, this buffer allocation step needs to be done, which is different from DNN models, and each new problem has a different shape and size.

The cost of using a scratchpad to do such buffer allocation amounts to exploring $>10^{80}$ choices, and here that needs to be performed for \textit{every new problem}. The use of scratchpads for this is impractical. \SpadName reduces this space to $10^2$, making it practical for every problem.

\insertWideFigure{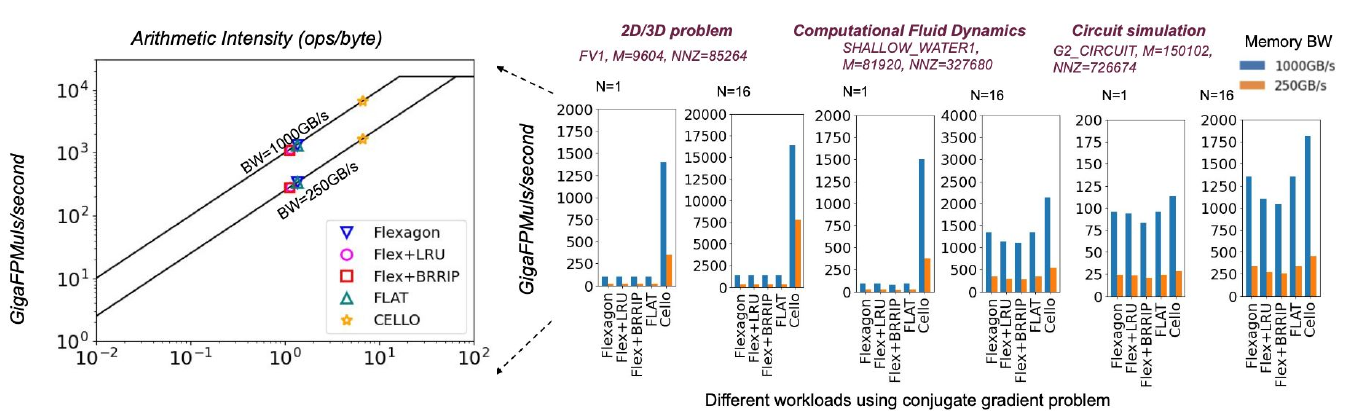}{Performance of accelerators (\autoref{table:dataflow_config}) for different problems that use CG algorithm (higher is better). First plot is shown on roofline. 
}

\insertFigure{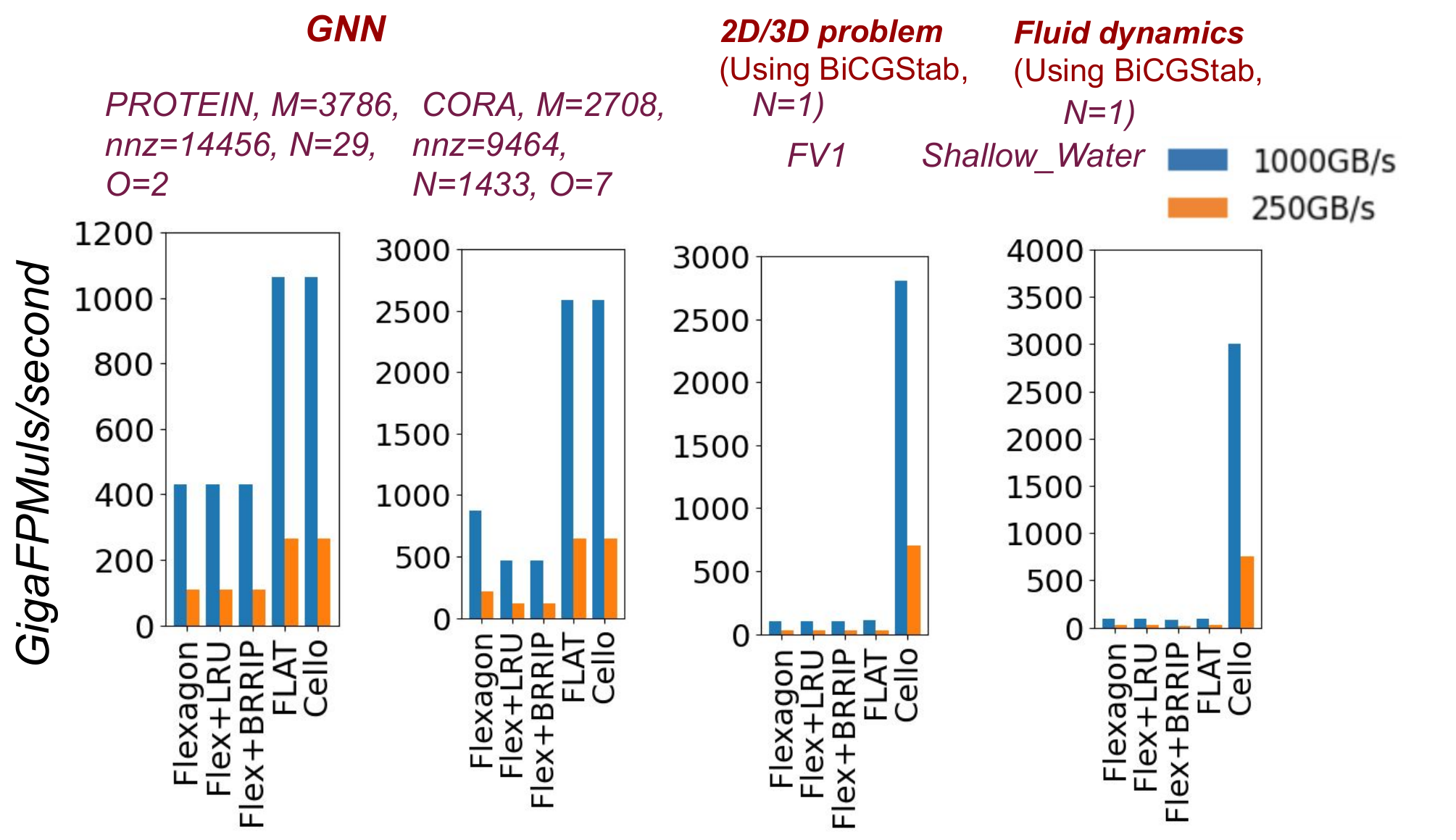}{Performance of accelerators (\autoref{table:dataflow_config}) for GNN and BiCGStab (higher is better).
}

\insertFigureScaled{energy}{Off-chip energy of schedules and buffer structures for different workloads (lower is better). Energy is geomeaned across different parameters or datasets in a workload.
}{1}

\insertFigure{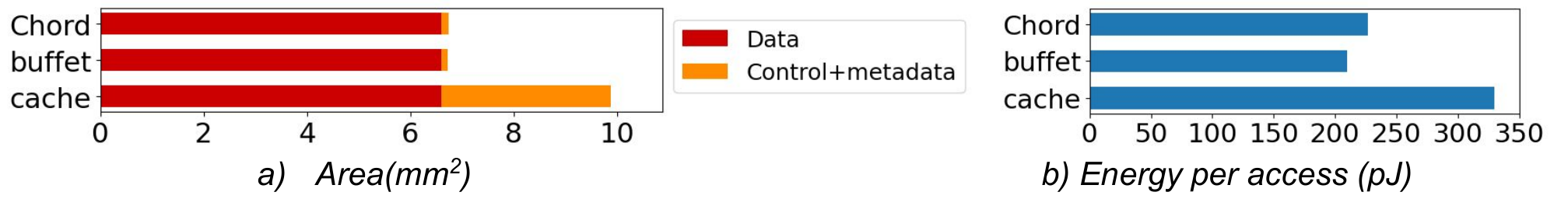}{a) Area and b) per access energy comparison (Lower is better).
}

\insertWideFigureScaled{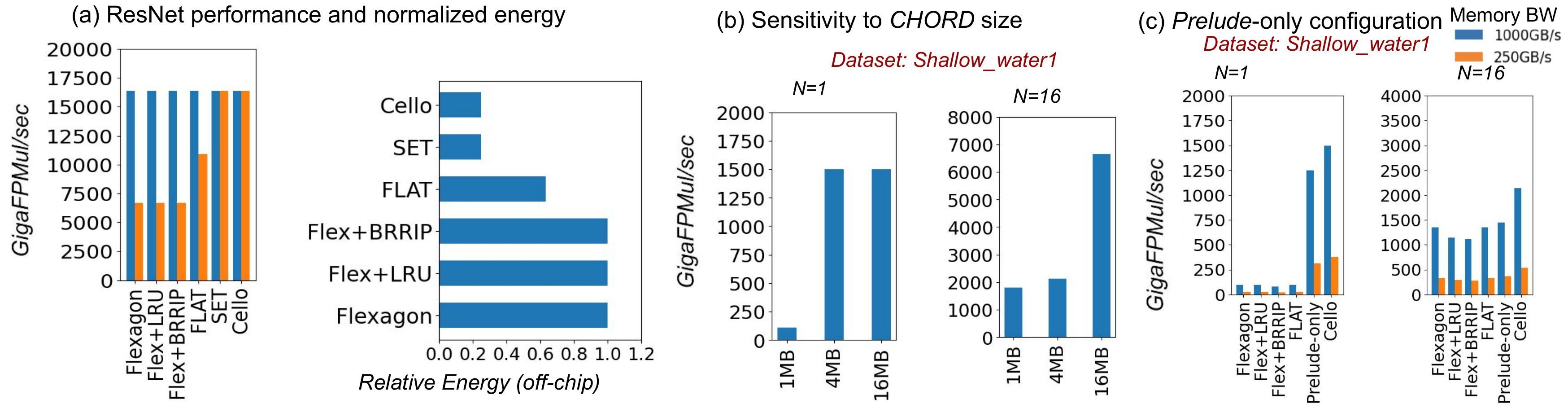}{All additional studies: (a) Performance and Energy of ResNet, with an additional baseline SET; (b)Variation of performance of \AccelName with \SpadName sizes on CG.; (c) \PolicyAnospace-only comparison against Flexagon, FLAT and \AccelName on CG.}{1}

\section{Evaluation}

\label{sec:eval}

\subsection{Methodology}

\subsubsection{Schedule and Buffer Baselines}
\label{sec:baseline}
We compare different combinations of schedules and buffer configurations described in detail in~\autoref{table:dataflow_config}. The Best intra-layer baseline dataflow corresponds to the oracle operation-by-operation dataflow, which only incurs the cold accesses, which corresponds to accessing each tensor once. In fact, such GEMMs are often able to achieve the best possible data reuse with memory accesses $MK\text{+}KN\text{+}MN$ since the tensor made of the non-dominant ranks can practically fit inside a portion of the Register File (RF) or parallelized across few PEs which means that the smaller tensor can just be accessed continuously from the RF while a tile of the large tensor is stationary. We keep the dominant rank in the outermost loop since the register file can store all of the small tensor. However, as discussed in~\autoref{sec:apps}, even the best individual op arithmetic intensity is low for skewed GEMMs.
We have chosen Flexagon's~\cite{flexagon} microarchitecture\footnote{\revision{Flexagon's microarchitecture is also capable of running dense GEMMs and SpMMs in addition to SpGEMMs}} to model the best intra-layer baseline because of its flexibility with the loop orders, which is required to achieve the best intra-layer baseline performance for SpMMs and GEMMs with contracted and uncontracted dominant ranks.
Note that the efficiency within the compute array does not matter significantly in this work since 
since stalls due to memory bandwidth dominate the delay. Therefore, we model the oracle op-by-op dataflow, which is the upper-bound for prior works (\ref{tables:related_score}) that use op-by-op execution. Similarly, FLAT configuration further considers adjacent pipelining capabilities in addition to to oracle op-by-op dataflow and is the oracle pipelined dataflow.

\subsubsection{Evaluation Framework}

We use LRU and BRRIP policy simulators to model set associative caches. We use OMEGA~\cite{garg2021understanding} and STONNE~\cite{STONNE21} to evaluate accesses for explicit baseline that emulates FLAT-like and Flexagon-like architectures.
To implement~\AccelNamenospace, we additionally, model writeback/hold and buffer occupancy behavior due to~\PolicyA and~\PolicyBnospace.  We model the area and access energy for caches and~\SpadName using CACTI~\cite{cacti}. We model off-chip energy based on the DRAM accesses.
\autoref{table:config} presents the hardware parameters.

\subsubsection{Workloads and Datasets}

\autoref{table:dataset} shows the HPC workloads and datasets.
We use the CG algorithm to model different problems - 2D/3D problem, computational fluid dynamics and circuit simulation.  
We obtain the sparse matrices for these
from Suitesparse~\cite{suitesparse}.
We also obtain GNN graphs from OMEGA~\cite{omega}.
We also evaluate \AccelName on ResNet in \autoref{sec:sens}.




\subsection{Main Results}


\subsubsection{Performance of Accelerators} - We compare the schedule and buffer-hierarchy configurations listed in~\autoref{table:dataflow_config} for workloads and datasets in~\autoref{table:dataset}. \autoref{fig:AI} and~\autoref{fig:AI2} show the performance across these workloads. \AccelName achieves 4x geomean speedup.


\textbf{Conjugate Gradient}: As~\autoref{fig:AI} shows, we observe that~\AccelName(\DataflowNamenospace+\SpadNamenospace) provides the highest throughput compared to the baselines evaluated. This is due to~\DataflowNamenospace's ability to identify pipelining and pipelining\_with\_writeback scenarios and~\SpadNamenospace's abililty to reuse intermediate tensors within the cache. Conjugate Gradient has operations where intermediate tensor always has a delayed downstream consumer. Therefore, works that only consider pipelining and parallel\_multicast are not beneficial here. LRU and BRRIP perform worse than best case schedule with explicit management.
Across the datasets, \AccelName varies in the amount of benefit. For smaller datasets, \DataflowName easily uses space inside the buffer. For larger working sets, \SpadName has better overall conflict miss management despite capacity limitations, thus allowing \AccelName to perform better.

\textbf{BiCGStab}: BiCGStab~\cite{bicgstab} is a different PDE solver algorithm, which can be used to solve the same PDE solver problems in~\autoref{table:dataset}. Like CG,~\AccelName outperforms the baseline for BiCGStab as well.

\textbf{Graph Neural Networks}: Within graph neural networks, \AccelName achieves the same performance as FLAT(\textit{Pipe+Exp}). The reason for the is that only tensor to be reused across operations in a GNN layer, is pipelineable without additional dependency. These configurations outperform Flexagon-like and its variants. For cora, due to the large feature map size, cache policies perform worse than Flexagon-like with explicit management.

\subsubsection{Off-chip Energy}

~\autoref{fig:energy} shows the workload-wise geomeaned off-chip energy, relative to \textit{BestIntra+Exp}. Our proposed \AccelName has the lowest energy for each workload, and the its reduced by 64\% to 83\%. Cache replacement policies have relatively higher energy compared to the explicit due to conflict misses. FLAT misses out on reuse along delayed dependencies. \AccelName achieves 4x geomean reduction in energy.

\subsubsection{Area and Energy of Buffer Structures}

\autoref{fig:areapower2}a and b show the area and energy comparisons of 4MB buffet, caches and~\SpadNamenospace. The area of buffets is the 6.72 mm\textsuperscript{2} as their controller adds only 2\% area overhead over data array~\cite{buffets}. Cache on the other hand has the total area of 9.87mm\textsuperscript{2}, with data array being 6.59mm\textsuperscript{2} and tag array being 1.85mm\textsuperscript{2}.~\SpadName has the area of 6.74mm\textsuperscript{2}.~\PolicyBnospace-index table requires 0.01x area compared to tag area in cache, to store information about 64 tensors across many operations. Energy per access is far greater for cache compared to buffets and~\SpadName. Infact, sizeable chunk of cache's energy comes from tag comparisons which ~\SpadName avoids.

\subsection{Additional Studies}
\label{sec:sens}

\subsubsection{Efficacy on DNN's with Complex DAGs: ResNet}

\autoref{fig:sens}(a) shows the performance and off-chip energy for all the configurations with ResNet. At 1TB/sec, ResNet is compute bound because the memory bandwidth is sufficient to saturate the compute. However, with memory bandwidth of 250 GB/s, minimum arithmetic intensity for compute bound changes from 16.384 ops/byte to 65.536 ops/byte. We also show the energy advantage of~\AccelName for ResNet. SET~\cite{isca-pip} takes care of delayed hold dependency (but not delayed writeback) and performs same as \AccelName on ResNet. However, it performs the same as FLAT and Flexagon on CG workload and is worse than \AccelName, since CG relies on delayed writeback dependency, which is not supported in SET.

\subsubsection{Sensitivity Study: SRAM Sizes}

\autoref{fig:sens}(b) shows the effect of reducing the~\SpadName capacity on \AccelNamenospace. In particular, \PolicyA is also affected by memory capacity. For larger N, the occupancy of the buffer for a tensor increases, which leads to higher occupancy causing the memory traffic to increase. With increasing size,~\SpadName is able to make replacement decisions more efficiently, even though its not too large for the workload at N=16. For N=1, the 4MB and 16MB SRAM are sufficiently large for the workload.

\subsubsection{\PolicyAnospace-only+Flexagon Configuration}

\autoref{fig:sens}(c) shows the CG results with additional \PolicyAnospace-only configuration to study the effect of \PolicyA alone. For CG, \PolicyA provides an advantage over "Flexagon" baseline and the "FLAT" baseline, since writeback support is more helpful than traditional pipelining support in case of CG. However, \PolicyB policy in \AccelName is effective in considering reuse frequency of a tensor, allows more frequent tensors to stay resident in \SpadNamenospace, which \PolicyAnospace-only does not consider, and hence \AccelName perfroms better. \PolicyAnospace-only configuration is closer to the baselines for N=16, while its closer to \AccelName with N=1. This is because \PolicyA benefits from tensor size relative to the on-chip SRAM.





\vspace{-1mm}



\section{Related Work}

\textbf{Scheduling tensor-algebra operations:} \autoref{tables:related_score} shows the related works on schedulers and their capabilities. Initial works~\cite{kwon2019understanding,timeloop,interstellar,kao2020gamma,cosa} propose schedule optimization search space, cost model or a search algorithm for a single tensor operation at a time. Prior works \cite{flat,garg2021understanding,genetic-pipeline,tileflow,FlashAttention,xlayer} propose a scheduling search space, cost model or novel schedules for pipelining between two operations. Some works~\cite{tangram,isca-pip} do map ResNet~\cite{resnet} demonstrating delayed hold capability which can be handled by pipeline buffer. However, pipeline buffer cannot be used in situations when there is a downstream consumer with a delayed\_writeback dependency. Thus, \DataflowName is more comprehensive than prior work.

\textbf{On-chip data orchestration}: 
~\autoref{tables:related_chord} shows the common buffer mechanisms. Buffets~\cite{buffets} is a recent work that uses explicit-decoupled data orchestration. Buffets are ideal for scheduling one operation at a time, however, explicitly allocating multiple intermediate tensors statically in an arbitrary DAG of einsums is a hard problem. Recent work Tailors~\cite{tailors} is another work on data orchestration targeting the problem of irregular tile sizes that spill over the buffer space. It employs a hybrid solution by automatically shrinking the element from the tail and replacing the new element by it in a coupled fashion. ~\SpadName is also a hybrid design-point. The policies implicitly orchestrate data elements in~\SpadNamenospace, while they obtain highly coarse-grained information (for example, tensor's global address range and tensor-level reuse distance and frequency) from the workload. 

\textbf{Conjugate Gradient accelerators:} Prior accelerators like Cerebras~\cite{rocki} and Azul~\cite{feldmann2024azul} use all-SRAM architectures with Giga-bytes large SRAMs to avoid DRAM accesses alltogether. In contrast \AccelName demonstrates speedup on traditional accelerators with few Mega-bytes of SRAM. Alrescha~\cite{asgari2020alrescha} is another accelerator, but it focuses on sparse operations alone, which have lower arithmetic intensity to begin with, while \AccelName lifts the upper limitation of arithmetic intensity by exctracting inter-operation reuse.

\section{Conclusion}
\label{sec:discussion}
\vspace{-1mm}
Ideally, tensor algebra applications would be made up of kernels that can be independently analyzed and optimized, then composed together into dependency graphs. In this paper we demonstrate that this individual operation view is too limited in scope: there is major opportunity to improve optimization by using an inter-operation viewpoint to increase arithmetic intensity, and that inter-operation pipelining fails to exploit reuse in the face of complex dependencies. Our proposed scheduler~\DataflowName identifies the complex reuse opportunities and the downstream tensors are reused by our proposed buffer structure~\SpadName at a tensor granularity. Our accelerator~\AccelName achieves 64\% to 83\% reduction in DRAM accesses compared to the state-of-the-art.

\section*{Acknowledgment}
Support for this work was in part provided through the ARIAA co-design center funded by the U.S. Department of Energy (DOE) Office of Science, Advanced Scientific Computing Research program. Sandia National Laboratories is a multimission laboratory managed and operated by National Technology and Engineering Solutions of Sandia, LLC., a  wholly owned subsidiary of Honeywell International, Inc., for the U.S. Department of Energy's National Nuclear Security Administration under contract DE-NA-0003525.

Part of this work was also supported by ACE, one of the seven centers in JUMP 2.0, a
Semiconductor Research Corporation (SRC) program sponsored by DARPA.

We also thank Axel Feldmann, Daniel Sanchez, Angshuman Parashar, and Nandeeka Nayak for their insightful feedback.


%

\clearpage
\bibliographystyle{IEEEtranS} 
\bibliography{refs}


\end{document}